\documentclass{elsart3p}

\usepackage{amssymb}
\usepackage{graphicx}
\usepackage{bm}

\begin{document}

\begin{frontmatter}


\title{(Ga,Mn)As on patterned GaAs(001) substrates: Growth and magnetotransport}

\author{W. Limmer\corauthref{limmer}},
\author{J. Daeubler},
\author{M. Glunk},
\author{T. Hummel},
\author{W. Schoch},
\author{R. Sauer}
\address{Abteilung Halbleiterphysik, Universit\"at Ulm, 89069 Ulm, Germany}
\corauth[limmer]{Corresponding author. Tel: +49-731-50-26130, Fax: +49-731-50-26108,
E-mail: wolfgang.limmer@uni-ulm.de}

\begin{abstract}
A new type of (Ga,Mn)As microstructures with laterally confined electronic and magnetic
properties has been realized by growing (Ga,Mn)As films on [1$\bar{1}$0]-oriented ridge
structures with (113)A sidewalls and (001) top layers prepared on GaAs(001) substrates.
The temperature- and field-dependent magnetotransport data of the overgrown structures
are compared with those obtained from planar reference samples revealing the coexistence
of electronic and magnetic properties specific for (001) and (113)A (Ga,Mn)As on a
single sample.
\end{abstract}

\end{frontmatter}

The dilute magnetic semiconductor (Ga,Mn)As, being compatible with conventional
semiconductor technology, is considered a potential candidate for spintronic
applications and has been intensely studied during the past few years \cite{Ohn98,Jun05}.
So far, (Ga,Mn)As layers were exclusively grown on planar substrates by low-temperature
molecular-beam epitaxy (LT MBE) where the structural, electronic, and magnetic properties
of the films have been shown to depend on the growth conditions and on the
crystallographic orientation of the substrate \cite{Omi01,Weg05,Wan05,Dae06}. Lateral
micro- or nanostructures for magnetotransport and magnetization studies were prepared
on homogeneous layers in a ''top-down'' procedure using various etching techniques. In
order to introduce an additional degree of freedom in the development of new functional
(Ga,Mn)As-based structures, we choose a completely different ''bottom-up'' approach.
We pick up a method widely used in the past to realize laterally confined semiconductor
microstructures in a single growth run, namely the growth of epilayers on patterned
substrates. More precisely, ridge structures exhibiting (001) top planes and (113)A
sidewalls were overgrown with (Ga,Mn)As films yielding (Ga,Mn)As layers with areas of
different substrate orientation on a single sample.

\begin{figure}
\hspace*{0.5cm}
\includegraphics{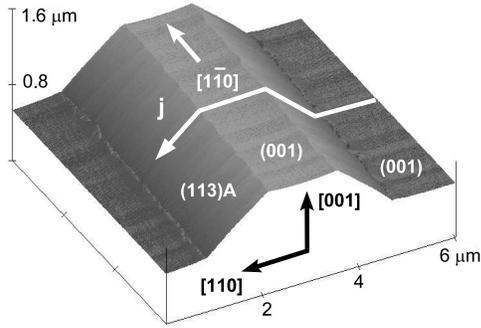}
\caption{\label{ridge_afm} AFM micrograph of an overgrown ridge structure with (113)A
sidewalls.}
\end{figure}

For this purpose, mesa stripes oriented along [1$\bar{1}$0] with (113)A sidewalls were
prepared on semi-insulating VGF GaAs(001) substrates by conventional photolithography
techniques and selective wet chemical etching. The sidewall orientation was controlled
by varying the composition of a NH$_4$OH:H$_2$O$_2$:H$_2$O solution. Etch depths between
0.2 and 0.6 $\mu$m were realized resulting in widths of the (113)A facets from 0.5 to
1.4 $\mu$m. The width of the (001) top plane was varied from 0 to 15 $\mu$m. After
removal of the photoresist the patterned substrates and planar (001) and (113)A
reference substrates were mounted with In on the same Mo holder. Then they were
introduced into a RIBER 32 MBE chamber equipped with a conventional Knudsen cell and a
hot-lip effusion cell providing the Ga and Mn fluxes, respectively. A valved arsenic
cracker cell was operated in the non-cracking mode to supply As$_4$ with a V/III flux
ratio of about 5. After thermal deoxidation under As$_4$ overpressure, a GaAs buffer
layer around 30 nm thick was deposited at a temperature of $T_s$ $\approx$ 580 $^\circ$C
(conventional substrate temperature for GaAs). Then the growth was interrupted and $T_s$
was lowered to $\sim$250 $^\circ$C. Finally, (Ga,Mn)As layers with Mn contents up to 5\%
and a thickness of 40 nm were grown.
Fig.~\ref{ridge_afm} shows as an example the surface topography of an overgrown
1.5-$\mu$m-broad ridge imaged by scanning atomic force microscopy (AFM). It reveals
smooth (001) and (113)A facets with sharp edges as well as a bump-like shape at the
(113)A/(001) facet transition due to the migration of Ga adatoms during the
high-temperature growth of the GaAs buffer layer \cite{Lim04}. No indication of the
presence of MnAs clusters is observable.

\begin{figure}
\hspace*{0.5cm}
\includegraphics{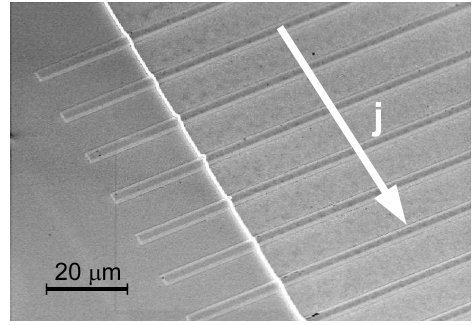}
\caption{\label{hallbar_sem} SEM micrograph of a Hall bar (section) carrying a series of
overgrown ridges. The edge of the Hall bar is seen as a white line.}
\end{figure}

The electronic and magnetic properties of the samples under study were investigated by
means of magnetotransport measurements, representing a powerful tool for the
characterization of ferromagnetic semiconductors. For this purpose, Hall bars were
prepared on the patterned samples and on the planar (001) and (113)A reference samples
with the current direction along [110] and [33$\bar{2}$], respectively. The Hall bars on
the patterned samples carry series of evenly spaced parallel ridges of equal dimension
oriented perpendicular to the current direction. A scanning electron microscopy (SEM)
micrograph showing a section of such a Hall bar is presented in Fig.~\ref{hallbar_sem}.

\begin{figure}
\hspace*{-0.5cm}
\includegraphics{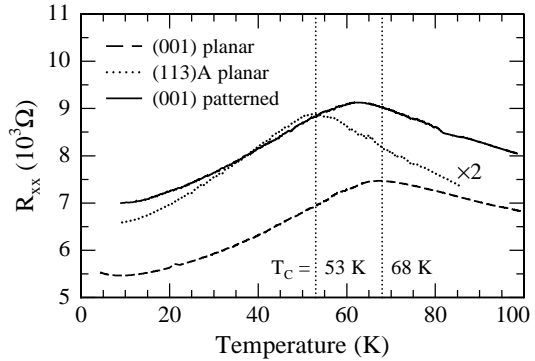}
\caption{\label{mt_temp} Estimate of the Curie temperatures $T_C$ from the peak positions
of the temperature-dependent resistances $R_{xx}$.}
\end{figure}

In the following, we focus on a sample with 1.5-$\mu$m-broad mesa stripes (see
Fig.~\ref{ridge_afm}) and a Mn content of 5\%. The Curie temperatures of the structured
layer and the corresponding reference layers were estimated from the temperature-dependent
longitudinal resistance $R_{xx}$ with an external magnetic field of 100 mT applied along
the [001] direction. From the peak positions of the measured curves depicted in
Fig.~\ref{mt_temp}, values of $\sim$68 K for the (001) and $\sim$53 K for the (113)A
reference layers are deduced. These values are consistent with the hole densities of
3$\times 10^{20}$ cm$^{-3}$ and 1.5$\times 10^{20}$ cm$^{-3}$, respectively, determined
from high-field (up to 14.5 T) magnetotransport measurements at 4.2 K. The curve recorded
from the patterned sample (solid line) exactly represents a superposition of the (001)
and (113)A curves (dashed and dotted lines, respectively) weighted according to the area
ratio of 82:18 between the (001) and (113)A planes on the Hall bar.

\begin{figure}
\hspace*{-0.5cm}
\includegraphics{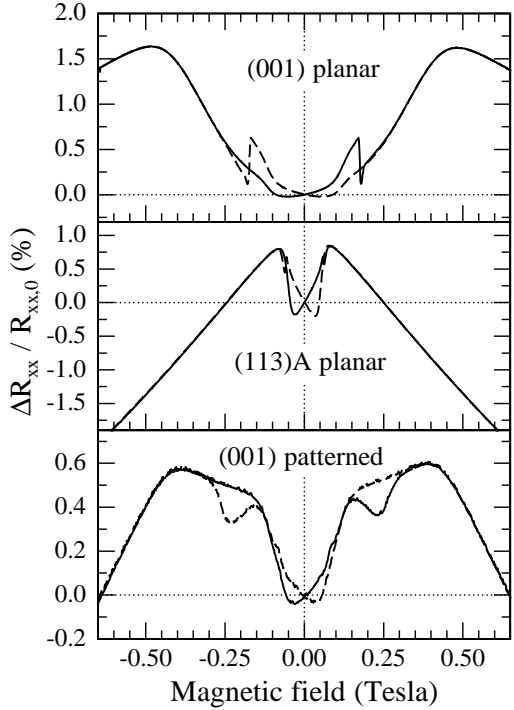}
\caption{\label{mt_field} Relative change of the longitudinal magnetoresistances at 4.2 K
measured as a function of magnetic field strength with the field $\bm{H}$ oriented along
[001].}
\end{figure}

Another, even more obvious evidence for the coexistence of (001)- and (113)A-specific
properties in the structured (Ga,Mn)As layer is obtained from the field dependence of
$R_{xx}$ measured at 4.2 K. In Fig.~\ref{mt_field} the relative changes
$\Delta R_{xx}/R_{xx,0}$ are plotted as a function of the magnetic field strength in
the range from -0.6 to 0.6 T for both, up-sweeps (solid lines) and down-sweeps (dashed
lines). For all samples the field $\bm{H}$ was applied along [001].
For sufficiently high magnetic fields the measured curves are dominated by a negative
magnetoresistance which has been ascribed to spin-disorder scattering \cite{Ohn98} and/or
suppression of weak localization \cite{Mat04}. The behaviour of $R_{xx}$ at low fields is
governed by an anisotropic magnetoresistance and by magnetic anisotropy. The longitudinal
and Hall resistances in (Ga,Mn)As are well known to sensitively depend on the orientation
of the magnetization $\bm{M}$ with respect to the current direction $\bm{j}$ and the
crystallographic axes \cite{Liu05,Goe05}. At zero magnetic field the magnetization is almost
parallel to the (001) plane, even in the case of the (113)A-oriented sample \cite{Lim06}.
With increasing field a reorientation of $\bm{M}$ along $\bm{H}$, i.e., along the [001]
direction takes place resulting in an increase of $R_{xx}$. Whereas in the (001) planar
sample the transition from anisotropic to negative magnetoresistance occurs at about
$\pm$0.5 T, the transition in the (113)A planar sample appears at $\pm$0.1 T. The patterned
sample clearly exhibits a superposition of both features confirming the simultaneous presence
of areas with (001)- and (113)A-specific properties. The jump in the curve of the (001)
planar sample at $\pm$0.18 T is due to magnetic anisotropy and arises from a sudden movement
of $\bm{M}$ caused by a discontinous displacement of its equilibrium orientation.

In summary, (Ga,Mn)As films of high crystalline quality were grown by LT MBE on
[1$\bar{1}$0]-oriented ridges with (113)A sidewalls prepared on GaAs(001) substrates.
Magnetotransport measurements clearly reveal the lateral coexistence of areas with (001)- and
(113)A-specific properties opening the prospect of a new degree of freedom in the development
of (Ga,Mn)As-based spintronic devices. (Ga,Mn)As microstructures realized in this way may be
easily overgrown with further layers without the need to interrupt the MBE growth.

\vspace{0.2cm}
This work was supported by the Deutsche Forschungsgemeinschaft, DFG Li 988/4.

\end{document}